\documentclass[
cpp,
a4paper,
fleqn,
finallayout
]{w-art}
\usepackage{times,cite,w-thm}
\usepackage{graphicx}
\usepackage{amsmath}
\usepackage{amsfonts}
\usepackage{amssymb}
\begin{document}
\title[Modeling and Simulation of IEDFs]{Modeling and simulation
of ion energy distribution functions in technological plasmas}
\author[T. Mussenbrock]{Thomas Mussenbrock}
\address{Institute of Theoretical Electrical Engineering,
Ruhr University Bochum, D-44780 Bochum, Germany}
\begin{abstract}
The highly advanced treatment of surfaces as etching and deposition
is mainly enabled by the extraordinary properties of
technological plasmas. The primary factors that influence these 
processes are the flux and the energy of various species, particularly ions,
that impinge the substrate surface. These features can be theoretically described
using the ion energy distribution function (IEDF). The article is intended to
summarize the fundamental concepts of modeling and simulation of IEDFs from
simplified models to self-consistent plasma simulations. Finally, concepts
for controlling the IEDF are discussed.
\end{abstract}
\maketitle

\section{Introduction}
It was almost ten years before the advent of plasma physics in
the beginning of the 1930th by Langmuir and Tonks
that it was claimed that energetic ions should play an important
role for the deposition of metals. In 1920, The Physical Society
of London and The Optical Society discussed ``The Making of
Reflecting Surfaces''. It was argued that ``the production of
mirrors by cathodic bombardment is not a new process''. Since
Sir William Grove, who is better known as one of the two fathers
of the fuel cell, discovered the phenomenon of sputtering in
1852, workers with vacuum tubes have noticed the bright deposit
in the vicinity of electrodes. It was found that this is more
or less due to the current passed through the tube and the
physical nature of the gas. They were convinced that the
mechanism of the process is doubtless that particles of the
metal are projected from the cathode by electric repulsion of
like charges. However, the discharges itself were crude apparatus
and the actual mechanism of sputtering was not understood.

Within the next fifty years -- until the 1960th -- the basic
concepts of today's low-temperature low-pressure plasma sources
and processes were developed. One important milestone was the
introduction of parallel plate capacitively coupled radio-frequency
discharges (CCPs) which were then widely used to generate
energetic ions for sputter etching and also sputter deposition.
Results from the first measurements of ion energy
distribution functions (IEDFs) were published by Er\"o in 1958.
He performed these measurements in the context of research on
a radio-frequency driven ion source of Thoneman type, which
is an inductively coupled plasma source \cite{Eroe1958}.
A typical IEDF from the measurements of Er\"o is depicted in
figure 1. It shows the typical bimodal shape due to the
radio-frequency modulation with a breadth of 55 eV. Another 
very important milestone which strongly influenced later research
activities on IEDFs has been published in 1963. Butler and Kino
studied the plasma sheath formation by radio-frequency fields
\cite{Butler1963}. In this context the role of the external circuit,
i.e., the blocking capacitor and the formation of a DC self-bias
voltage were discussed for the first time. 

\begin{figure}
\includegraphics[width=0.7\linewidth]{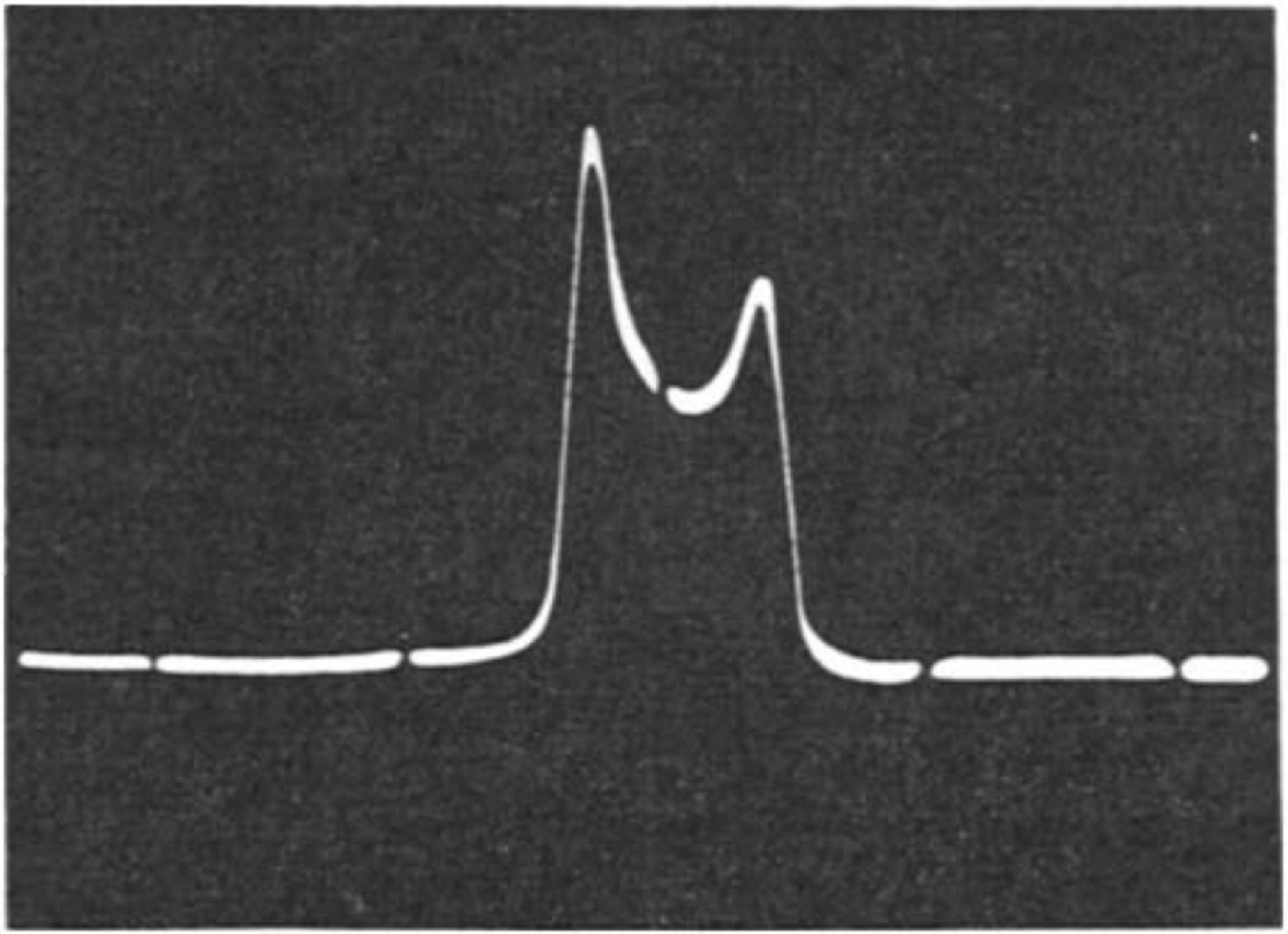}
\caption{Energy spectrum at 3 keV from Er\"o \cite{Eroe1958}.
The figure shows the characteristic bimodal shape due to the
radio-frequency modulation. The distance between the two maxima
corresponds to 55 eV.}
\end{figure}

In the late 1960th and the early 1970th CCPs were mainly applied
for plasma etching or sputtering. Plasma etching was more or less 
isotropic and combined volatile products and low ion energies.
On the other hand sputtering combined nonvolatile products and
high ion energies. It was actually the time that the need for
volatile etch product and anisotropic etch (high ion energies)
was recognized. In 1974 Hosokawa et al. were the first who combined
volatile products and high ion energies driven by their interest
to realize faster plasma etching \cite{Hosokawa1974}. They found
that a fluorine plasma of a CCP can be a very useful source of
both energetic ion fluxes and reactive neutral species when the
substrate is placed on the powered electrode. It is of course
due to the sheath which is formed between the substrate and the
plasma and in which the highly mobile electrons are absent to form
a strong electric field allowing an energetic flux of positively charged particles
to the wall. The quite sharp potential gradient causes the positive
ion flux to maintain a narrow angular distribution at the wafer
which ultimately leads to anistropic etching.

At least since Coburn and Winters in 1979 published their famous 
work on plasma-assisted surface treatment, low-temperature
low-pressure plasmas began to revolutionize the materials
processing scene \cite{Coburn1979}. Coburn and Winters found
that a synergistic effect appears when combining an argon ion
beam with a fluorine atom flux from a hot XeF$_2$ gas. They showed
in their famous figure (figure 2) that due to the combined
fluxes the etch rate is an order of magnitude larger than that
produced by either individual flux. It was also found by Coburn
and Winters that the required activation of the process by the ion
flux allowed the anisotropic nature of the etching process to be
retained. This actually could be named the advent of reactive ion
etching (RIE), or even the advent of plasma-based materials
processing. Driven by the quest of the semiconductor industry
to further satisfy Moore's law, plasma processing plays nowadays
a crucial role for the fabrication of most of the available
hi-tech products.

\begin{figure}
\includegraphics[width=0.7\linewidth]{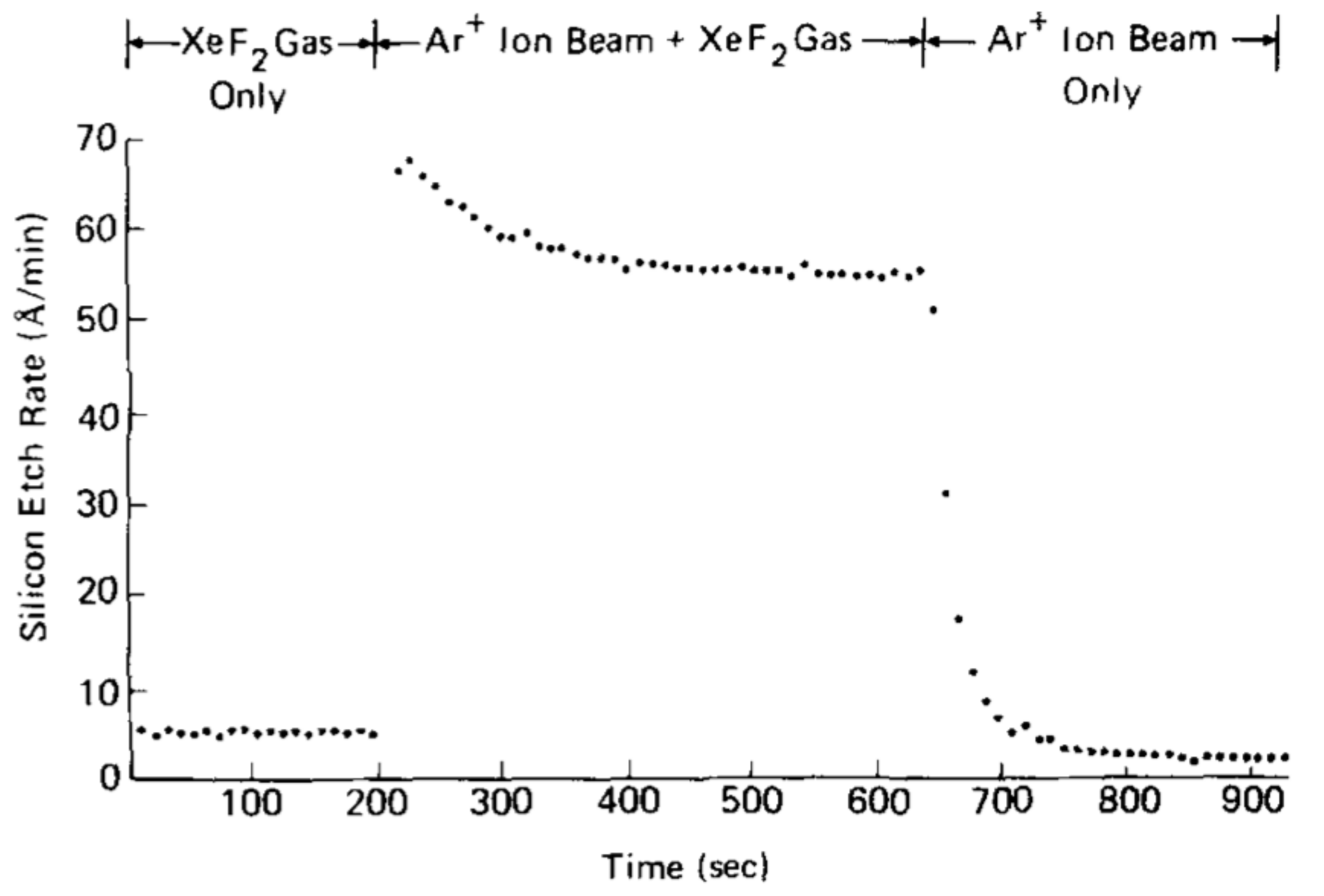}
\caption{Ion-assisted etching of a silicon substrate using
either argon ions or xenon difluorid gas, or both at the same
time from Coburn and Winters \cite{Coburn1979}.}
\end{figure}

In order to allow for anisotropic etching at high etch rate,
modern plasma sources are operated at high plasma densities and
low gas pressures. The sheath voltages are often quite small to
prevent ion induced damaging of the structured substrates. In 
this regime the mean width of the sheath is much smaller than
the mean free path of the ions. Therefore the ions traverse
the sheath without any collision and the spread of related IEDFs
and IADFs (ion angular distribution functions) is minimal. At
higher pressures the number of collisions of ions and neutrals
of the background become significant. The result is a very
complex and broad shape of the IEDF in contrast to the beam-like
distribution at low pressures.

Over the last decades a large number of modeling and simulation
approaches to low-temperature low-pressure plasmas, to their sheaths,
and inherently to IEDFs have been published. (An excellent review has
been povided by Kawamura et al. in 1999 \cite{Kawamura1999}.)
Actually, it is impossible to explicitly acknowledge all activities
in this field. That is why this manuscript makes no claim to be complete.
In contrast, this paper is intended to
summarize the fundamental concepts of modeling and simulation of IEDFs from
simplified models to self-consistent plasma simulations. It is therefore
organized as follows: After this introductory part, the basic simplified
models for calculating IEDFs are revisited. It will be argued that the
dynamics of sheath play a crucial role for the
characteristics of the IEDF. Therefore, the Brinkmann sheath model is
briefly discussed. The following fourth chapter deals with selfconsistent
plasma simulations in the context IEDF calculations. Finally, well-established
and new concepts for controlling the IEDF are discussed.

\section{Simplified models}

Analytical models for calculating IEDFs are often
based on the simplifying assumption of a collisionless sheath.
The  mean free path of the ions is assumed to be much smaller than
the mean sheath width. In this context important parameters
are the ratio between the ion transit time $\tau_{\it ion}$ 
(the time for an ion
traverse through the sheath) and the time constant of 
the frequency of the sheath modulation $\tau_{\it rf}$, which is nearly
the time constant of the driving radio-frequency. Assuming a Child-Langmuir sheath
with a sheath potential $V_{s}\propto x^{4/3}$, one obtains
\begin{gather}
\frac{\tau_{\it ion}}{\tau_{\it rf}}= \frac{3 \bar s
\omega_{\it rf}}{2 \pi}\sqrt{\frac{m_i}{2 e \bar V_s}},
\end{gather}
with $\bar s$, $\bar V_s$, and $m_i$ being the mean sheath
width, the mean sheath voltage, and the ion mass, respectively.

One may distinguish between two regimes: The high-frequency
regime where $\tau_{\it ion}/\tau_{\it rf}\gg 1$ and the
low-frequency regime where $\tau_{\it ion}/\tau_{\it rf}\ll 1$.
In the latter regime the ions traverse the sheath within a very
short time, which can be only a very small fraction of one
low-frequency period. Therefore the ions experience the
instantaneous electric field of the sheath. The IEDF at the
electrode is strongly depending on the phase of the electric
field at the instant time when the ions arrive at the sheath edge.
In the low-frequency regime the IEDF has more or less the bimodal
shape, similar to the IEDF which is displayed in figure 1. The bimodal shape is
characterized by the distance between the two maxima $\Delta E$.
The maxima of the IEDF correspond with the minimum and the 
maximum of the sheath voltage, respectively. In the high-frequency
regime the ions traverse through the sheath within in a large
number of radio-frequency cycles. The ions experience more or
less the time-averaged electric field rather than the instantaneous
electric field. In this regime one obtains a narrow, beam-like
IEDF. It has been shown that for the high-frequency regime
the breadth of the IEDF is directly proportional to the ratio 
$\tau_{\it ion}/\tau_{\it rf}$. If $\tau_{\it ion}$ increases
for fixed $\tau_{\it rf}$,  $\Delta E$ becomes smaller. For
a very large ion transit time the bimodal IEDF becomes a
monoenergetic IEDF.

The first analytical calculation of the IEDF at a substrate was
proposed by Benoit-Catin and Bernard for the collisionless
high-frequency regime \cite{Benoit1968}. The model is based on 
a number of simplifying assumptions: In the frame of a
Child-Langmuir sheath model the sheath width is assumed
to be constant and the electric field is assumed to be uniform.
The temporal modulation of the sheath voltage is harmonic,
i.e., $V_s=\bar V_s + \hat V_s \sin \omega_{\it rf} t$.
Additionally, it is assumed that the ions start with zero
initial velocity at the sheath edge. A straight forward analysis
of Poisson's equation coupled with Newton's law of motion for
a charged particle in an electric field yields an expression
for the IEDF $f(E)$ and the energy spread $\Delta E$, respectively,
\begin{gather}
f(E)= \displaystyle\frac{ 2 n_t / \displaystyle \omega_{\it rf}
\Delta E}{\sqrt{1 -\displaystyle\frac{4}{\Delta E^2} 
\left(E - e \bar V_s \right)^2}}
\end{gather}
and
\begin{gather}
\Delta E = \frac{8 e \tilde V_s}{3 \bar s \omega_{\it rf}}
\sqrt{\frac{2 e \bar V_s}{M}}.
\end{gather}
$f(E)$ shows the typical bimodal shape with two maxima symmetrically
localized about $e \bar V_s$. The energy spread $\Delta E$ is
proportional to  $\tau_{\it rf}/\tau_{\it ion}$. With increasing
frequency or increasing ion mass, $\Delta E$ becomes smaller. Ultimately,
the two peaks approach to each other and become one monoenergetic peak.
Due to the monoenergetic initial velocity distribution one obtains
singular behavior of $f(E)$ at
\begin{gather}
E = e\bar V_s \pm \frac{\Delta E}{2}.
\end{gather}

\begin{figure}
\includegraphics[width=0.7\linewidth]{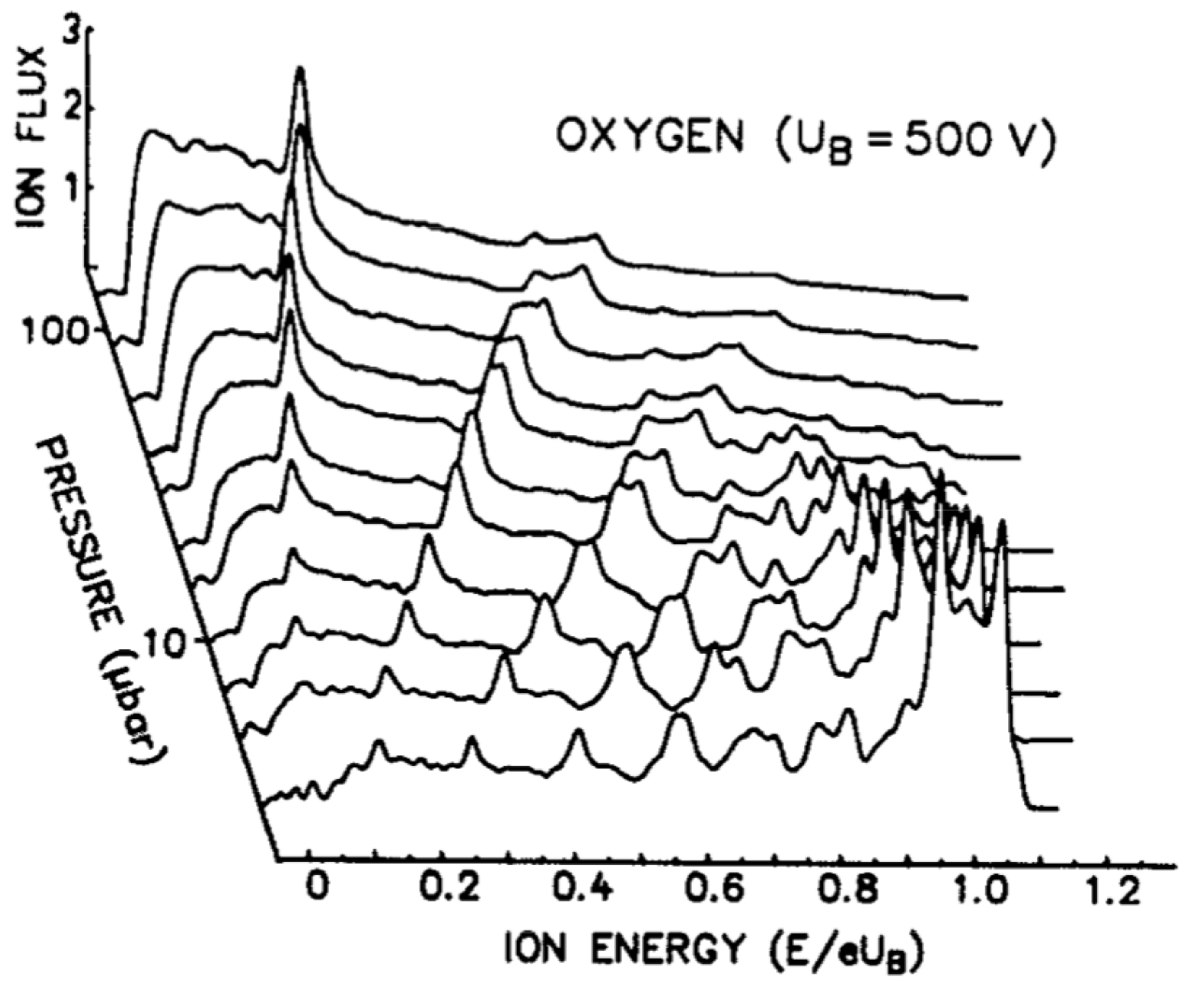}
\caption{Ion energy distributions measured in oxygen
discharges at different pressures from  Wild and Koidl
\cite{Wild1991}}
\end{figure}

The effect of collisions of ions with the neutrals of the background
gas within the sheath was first studied experimentally as well as
theoretically by Wild and Koidl \cite{Wild1991}. Figure 3 shows a
series of measured IEDFs in oxygen discharges at different gas
pressures. One can observe a number of peaks and double peaks whose
occurrence strongly depend on the pressure. At low pressures the
IEDF shows mainly the well-known bimodal structure at high energies
and a number of less significant peaks for lower energies. This
structure is due to ions which have traversed the sheath
without collisions. With increasing pressures, the bimodal
shape successively disappears and the average ion energy decreases
significantly.

The theoretical model proposed by Wild and Koidl is able to
resolve for the complex multi-peak structure of the IEDFs.
It is based on similar simplifying assumptions as the
collisionless model of Benoit-Catin and Bernard. All assumptions
are well justified and can be applied to the collisional case.
Of course, the most important issue in this context is the
consideration of collisions of ions with neutrals during their
traverse of the sheath. Of all ion collision processes within
the sheath, the charge exchange collision between the ions and
the neutrals have the largest cross-section. Thus, charge exchange
processes play the major role and are exclusively allowed for.
It can be found that even at low pressures the charge exchange
process is quite probable and dominant. It should not be neglected
when studying the ion transport through the sheath. 

\begin{figure}
\includegraphics[width=0.7\linewidth]{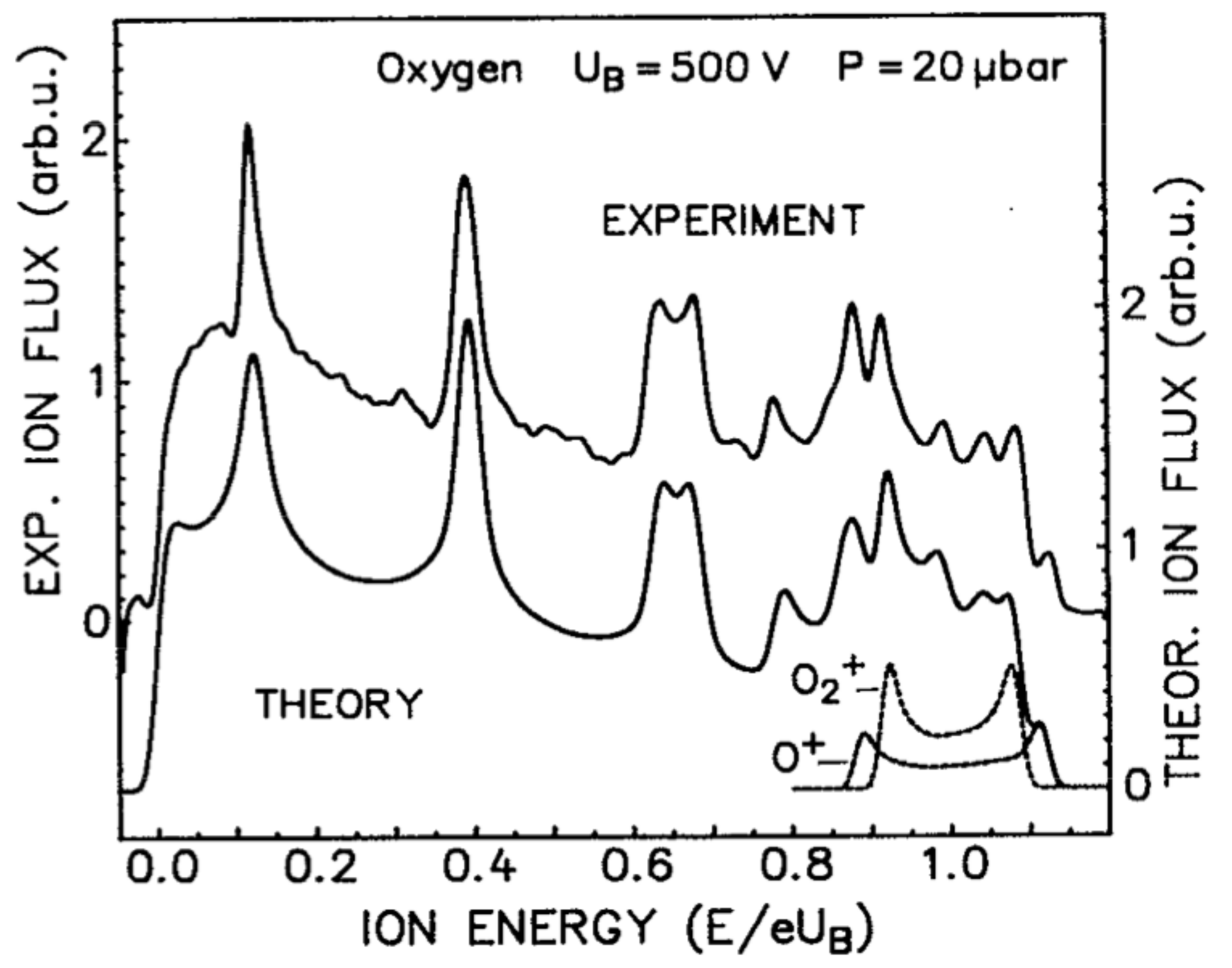}
\caption{Experimental (upper curve) and theoretical (lower curve)
energy distribution of the ions in an oxygen discharge from Wild
and Koidl \cite{Wild1991}. The small dashed lines in the lower right
corner represent the contribution of the unscattered ions to the
theoretical IEDF.}
\end{figure}

The charge exchange collision of an energetic ion with a
thermalized neutral particle produces an energetic neutral
and a thermalized ion under conservation of the total ion flux.
The assumption is that a thermalized ion is generated at the
position $p_0$.
The probability that this particular ion reaches the substrate
is given by $\exp \left[\alpha(1-p_0)\right]$ with $\alpha$
being the ratio between the mean sheath width and the ion mean free path
within the sheath. The energy of the ions that impinge on the
substrate is therefore a function of the position of the collision
$p_0$ and the instantaneous phase of the electric field $\phi_0$
at the instant of time of the collision. If ions traverse the sheath
without any collision, the energy is a function of the initial
velocity $v_0$ at the sheath edge and the instantaneous phase
$\phi_0$ of the electric field. For a constant collision cross
section one arrives at
\begin{gather}
f(E)dE \propto \int \text{e}^{-\alpha (1-p_0)}dp_0 d\phi_0 +
\text{e}^{-\alpha}\int f(v_0)dv_0d\phi_0,
\end{gather}
which is an expression for the IEDF allowing for charge exchange
collisions of ions within the sheath. Here $f(v_0)$ is the
distribution of the initial velocity of ions at the sheath edge.
The first part of this expression represents the contribution
of ions which are produced due to a charge exchange process
within the sheath. The second term is the contribution of ions
which traverse the sheath without any collision. It can be shown
that the initial ion velocity distribution at the sheath edge
has almost no influence on the IEDF at the substrate. Therefore
$f(v_0)$ is often assumed to be a Boltzmann distribution. Figure 4
shows both experimental and theoretical results for IEDFs in an
oxygen discharge. Whereas the upper curve shows the experimental
result, the lower curve is the result from the model. The small
dashed lines in the lower right corner represent the contribution
of the unscattered ions to the theoretical IEDF. One can find
that the results are in almost perfect agreement. In the 
displayed case of an oxygen discharge one can observe a shoulder
in the experimental results for high energies which can not be 
resolved when assuming only singly ionized oxygen molecules O$_2^+$.
To account for this structure, the contribution of ionized oxygen
atoms O$^+$ or double ionized oxygen molecules O$_2^{2+}$ have
to be included into the model.

\section{Self-constistent sheath models}

As already mentioned in the previous chapters the most important
parameter -- besides the gas mixture and gas pressure -- is
the electric field within the sheath region. To calculate the
spatial and temporal dynamics of the sheath (and  inherently
the electric field) a large number of stand-alone sheath
models have been proposed, e.g., by Lieberman, Biehler, Riemann,
or Godyak \cite{Lieberman1988, Lieberman1989, Biehler1989,
Riemann1989, Godyak1990, Godyak1993}. All models are more
or less able to resolve for the important
phenomena. A very elegant and powerful model has been developed by
Brinkmann \cite{Brinkmann2007, Brinkmann2009, Brinkmann2011}.
The model is able to treat an arbitrary number of positive
and negative ions and is therefore appropriate
in the context of technological applications.
Brinkmann's model is characterized
by a smooth transition from unipolarity (in the sheath) to
ambipolarity (in the bulk plasma), instead of a step-like transition. 
Capacitively driven sheaths exhibit a smooth transition,
which is expanded by the radio-frequency modulation
and smoothed by thermal effects which are namely the finiteness of
the electron temperature and the Debye length. Sheath models
which neglect thermal effects are characterized by a step in the
spatial electron density distribution. These so-called step models
are restricted to strongly modulated high voltage sheaths with
$V_{RF} \gg T_e/e$ and fail when this condition is not met.
To overcome this problem, Brinkmann proposed an improved model
of the sheath-bulk transition, which takes into account both
modulation and thermal effects. Based on an asymptotic
solution of the Boltzmann-Poisson equation one can derive an approximate
algebraic expressions for the phase-resolved electrical field
and electron density in radio-frequency driven sheaths. 
Under the assumption that the modulation of the sheath 
is periodic with $\omega_{RF} \gg \omega_{pi}$, also the phase
averages of the electric field  and the electron density can
be explicitly expressed in closed form. These results -- together
referred to as the Advanced Algebraic Approximation (AAA) -- make 
it possible to formulate efficient and accurate models for radio
frequency driven boundary sheaths for all ratios $e V_{RF}/Te$
and for an arbitrary number of positive and negative ions.

\begin{figure}
\includegraphics[width=0.7\linewidth]{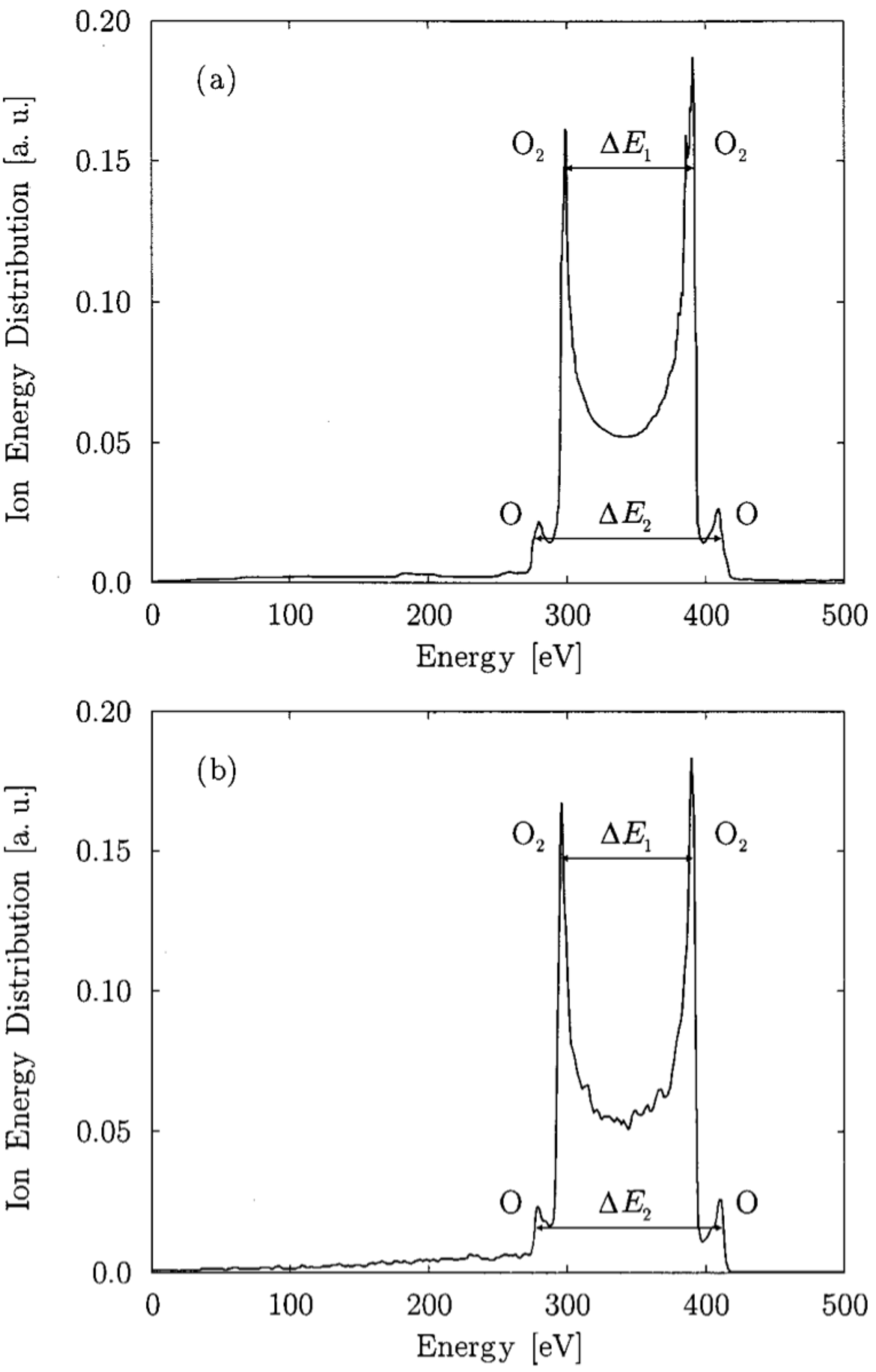}
\caption{Calculated and measured IEDFs of an oxygen discharge
at 3 mTorr from Kratzer et al. \cite{Kratzer2001}. The
experimental result obtained by Kuypers and Hopman is displayed
above \cite{Kuypers1990}. The corresponding simulation result
is given below.}
\end{figure}

Kratzer et al. applied the AAA to self-consistently calculate
the electric field for a number of different plasma parameters
and gas mixtures \cite{Kratzer2001}. The electric field has been
then used in a
Monte Carlo module in order to calculate the ion trajectories
within the sheath. The collisions of the ions with the neutrals of
the background gas are treated in the frame of the null collision
method. The related IEDFs from Brinkmann's sheath model have been
impressively validated. Figure 5 shows both numerical and experimental
results for IEDFs in a radio-frequency oxygen discharge at a gas
pressure of 3 mTorr. The experimental results were originally taken from
Kuypers and Hopman \cite{Kuypers1990}. To compare the experimental
data with the simulation results, the latter have been smoothed with a Gaussian
smoothing filter of 3 eV FWHM which was roughly the average resolution
in the experimental data. In both results one can identify two distinct
double peak structures which are due to the two positive oxygen ion
species O$_2^+$ and O$^+$. Whereas the inner peaks belong to O$_2^+$,
the species with the higher mass, the outer lying peaks belong to the
lighter ions O$^+$. In accordance with the theory described above
the peak separation is inversely proportional to the square root
of the ion masses
\begin{gather}
\frac{\Delta E_1}{\Delta E_2}\propto\sqrt\frac{m_2}{m_1}
\end{gather}
The calculated IEDF by means of Brinkmann's AAA displayed in figure 5
reproduce the experimental measurements in great detail. Of course, the
assumption that the ion plasma frequency is much larger than modulation
frequency of the sheath is a limitation of the AAA in its current stage.
However, the remedy of this drawback is a topic of ongoing research.

\section{Self-consistent discharge models}

Although stand-alone sheath models are able to provide meaningful results
for a wide range of plasma parameters, all these models have
an intrinsic drawback: They are not fully self-consistent in the
sense that the ion flux and energy at the sheath edge has to be
assumed. This drawback can be remedied by using fully self-consistent
discharge models. A large number of such models, which are based on
different approaches, are available. The main difference is manifested
in the fact, whether the approach is fluid dynamic, kinetic, or hybrid
(which means both fluid and kinetic).

\begin{figure}
\sidecaption
\includegraphics[height=1\textheight]{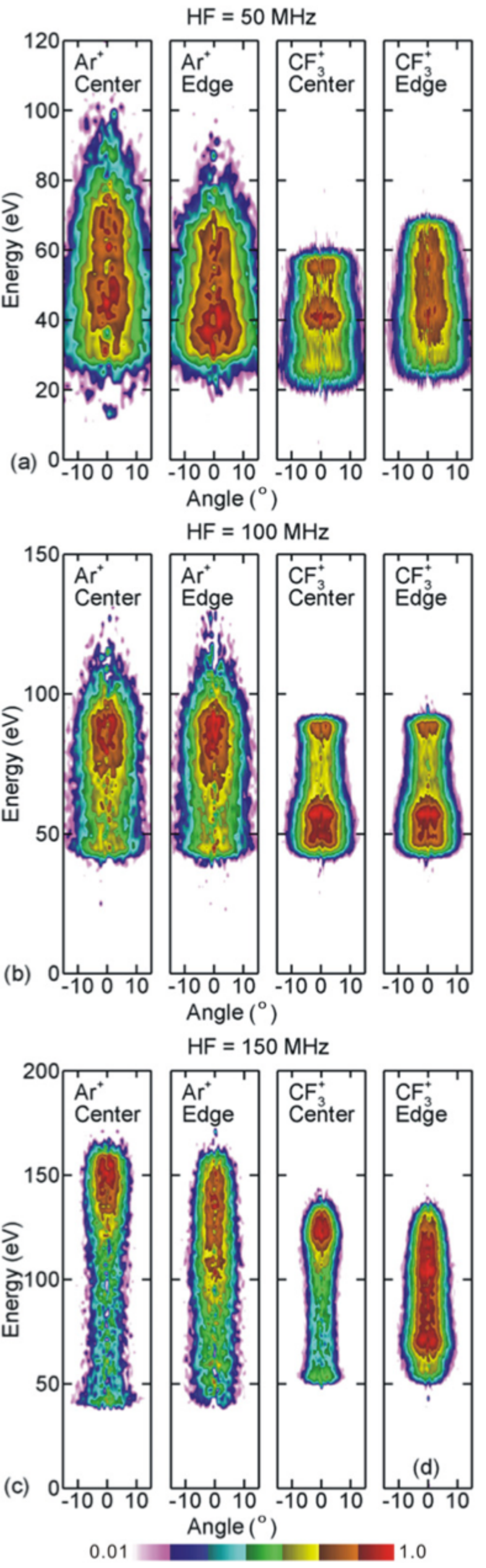}
\caption{IEDFs of different ion species
in an Ar/CF$_4$ plasma obtained by using HPEM for different
dual-frequency excitation from Yang and Kushner \cite{Yang2010}}.
\end{figure}

A very prominent and elaborated hybrid model is the Hybrid Plasma
Equipment Model (HPEM) which has been implemented over the last 30 years
by Kushner and co-workers \cite{Kushner2009}. It is devoted to
provide a simulation tool for low-temperature low-pressure plasma
systems, from external circuits to the feature length scale in order
to simulate plasma based surface processes. A huge number of
researcher and applicants from the industry have successfully
applied HPEM, not only to understand the basic plasma physics but also
to optimize industrial plasma processes. Similar to other available
hybrid simulation models, HPEM consists of a number of distinct
modules. Basically, it self-consistently solves continuity, momentum,
and energy equations for neutrals and ions, continuity equations
for electrons, and Maxwell's equations. It utilizes an electron
Monte Carlo simulation to obtain the electron energy distribution
functions of bulk electrons and secondary electrons emitted from
surfaces. The equations are integrated in time over many radio-frequency
cycles to obtain a periodic steady state. During the last iteration,
the converged electric fields and source functions for ions and neutral
are recorded as a function of position and phase in the radio-frequency
cycle. With these values, the energy and angular distributions of ions
and neutrals incident on the substrate are obtained using an adapted
plasma chemistry module based on a Monte Carlo scheme \cite{Lu2001}.
The IEDFs can then be
used to perform simulations of the actual plasma processes on the
length scale of the feature, e.g., the etching or deposition of materials. 

Figure 6 shows typical HPEM results for IEDFs of different ion species
in an Ar/CF$_4$ plasma \cite{Yang2010}.
The IEDFs of Ar and CF$_3$ ions as functions of the angle of impact
on the substrate
are displayed for a dual frequency excitation of 10 MHz/50 MHz,
10 MHz/100 MHz, and 10 MHz/150 MHz (see next section). One can observe
an increasing radial dependence of the maximum of the ion energy as the
high-frequency increases. This phenomenon can be explained by 
non-uniform plasma distributions at 50 and 150 MHz due to electromagnetic
effects as the standing wave effect and the skin effect \cite{Lieberman2002,
Chabert2007, Mussenbrock2008, Lee2008}. 

\begin{figure}
\includegraphics[width=0.7\linewidth]{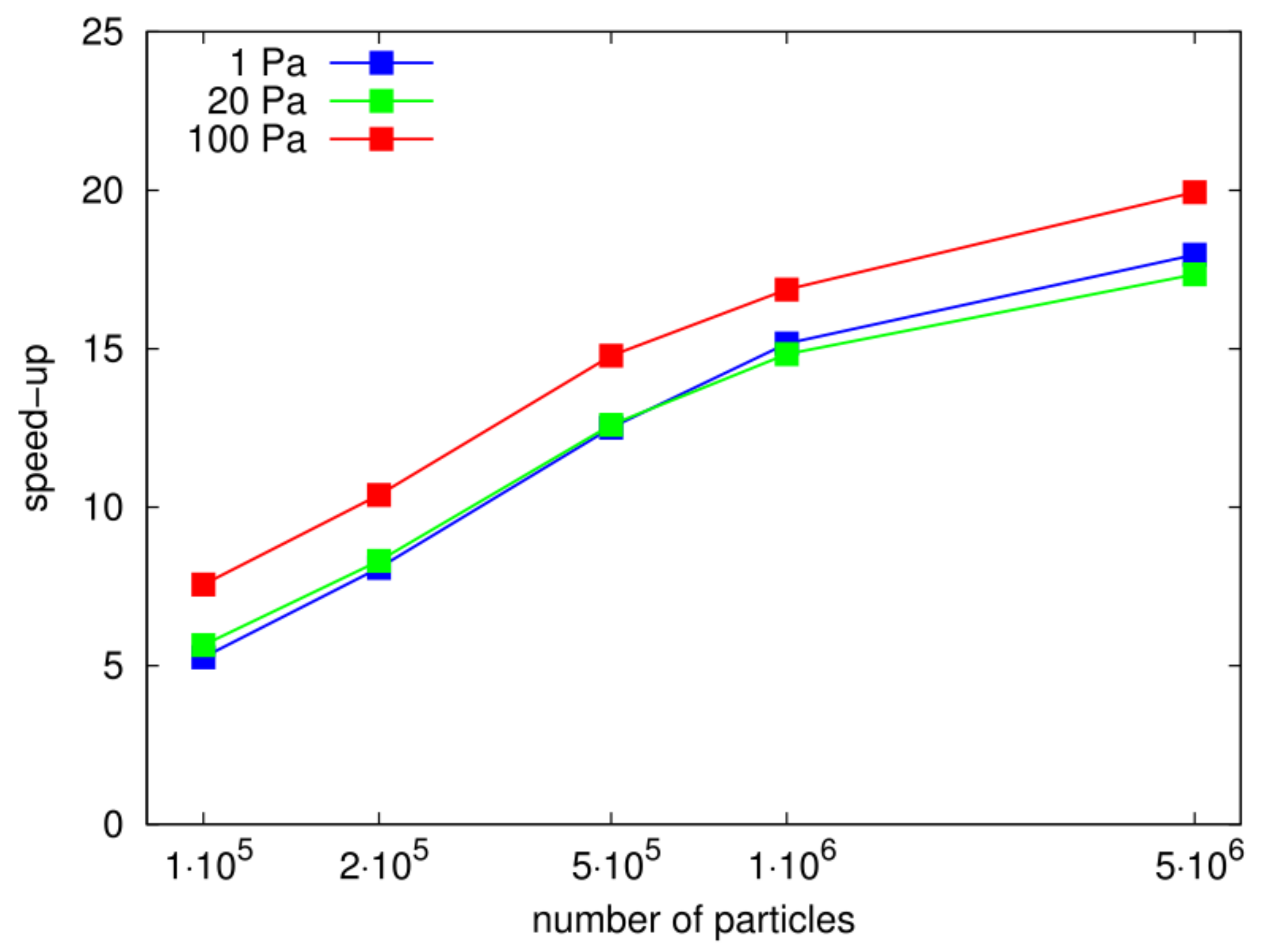}
\caption{Speed-up of the fine-sorted PIC algorithm, compared to a
CPU (single core) approach as a function of the number of
superparticles for different pressures from Mertmann et
al. \cite{Mertmann2011}.}
\end{figure}

However, although hybrid models provide very convincing
results, the most straightforward
way to simulate plasmas is the particle-in-cell (PIC) method
pioneered among others by Birdsall \cite{Birdsall1991} and firstly applied to
low-temperature plasmas by Vender and Boswell \cite{Vender1990}. They set up
a PIC model allowing for charge exchange collisions in the frame of
the Monte Carlo (MC) scheme of a 1D parallel plate radio-frequency
discharge in hydrogen. It was intended to study the IEDF at the
electrode for a number ofdifferent parameters.
The results confirmed both, the analytical results for the
simplified collisionless model proposed by Benoit-Catin and the
numerical results performed by Koidl and Wild using the collisional
model. It was Surendra
and Graves who first implemented a collision model included elastic
and ionizing electron neutral collisions in addition to charge
exchange collisions of ions \cite{Surendra1991}. They conducted simulations of
radio-frequency discharges in Helium which also confirmed the 
experimental results of Koidl and Wild.

Since the basic concept of the particle in cell method coupled
to Monte Carlo collisions is quite simple, a large number of
researchers and research groups implemented their own PIC code
in order to have their own numerical ``plasma experiment''.
The advantage of the PIC concept is that the fields and the
energy distribution can be obtained from first principles.
There is no need for simplifying assumptions for the fields
or the distributions. However, the advantage comes along with
a significant disadvantage. The PIC/MC scheme tends to be
computationally very expensive compared to other numerical
simulation methods. The good news is that modern computers
have become very fast and very large memories are affordable.
Particularly, modern graphics cards (or graphics processing
units GPUs), originally introduced to make computer games
more realistic and more fast, provide a
very strong computational infrastructure. Up to 1.000 cores on
one GPU are available. It has been shown that PIC/MC codes 
in running parallel on standard GPUs can easily reach  a speed-up
of nearly a factor of 80 compared to the serial PIC code for CPUs.
A second inherent advantage is that a very large number of particles
(up to a few 10 millions) can be simply tracked without losing
significant performance.

However, despite all technical advances which have been recently made
in the context of PIC/MC, there is still lack of knowledge about collision
cross section which are needed to implement reasonable collision processes
in the frame of the Monte Carlo scheme. If one is interested in more complex 
gas mixtures, e.g., which are used for plasma surface treatment, one would
probably use fluid models or hybrid models.
 
\section{Control of IEDFs}

The ability of modeling and simulation the IEDF in
technological plasmas forces the quest to actually
control the energy and the flux of the ions impinging
on the substrate in order to directly control plasma surface
interactions. A number of methods which are intended to 
control (or at least to adjust) the features of the
IEDF have been developed, or are still matter of ongoing
research.

\begin{figure}
\includegraphics[width=\linewidth]{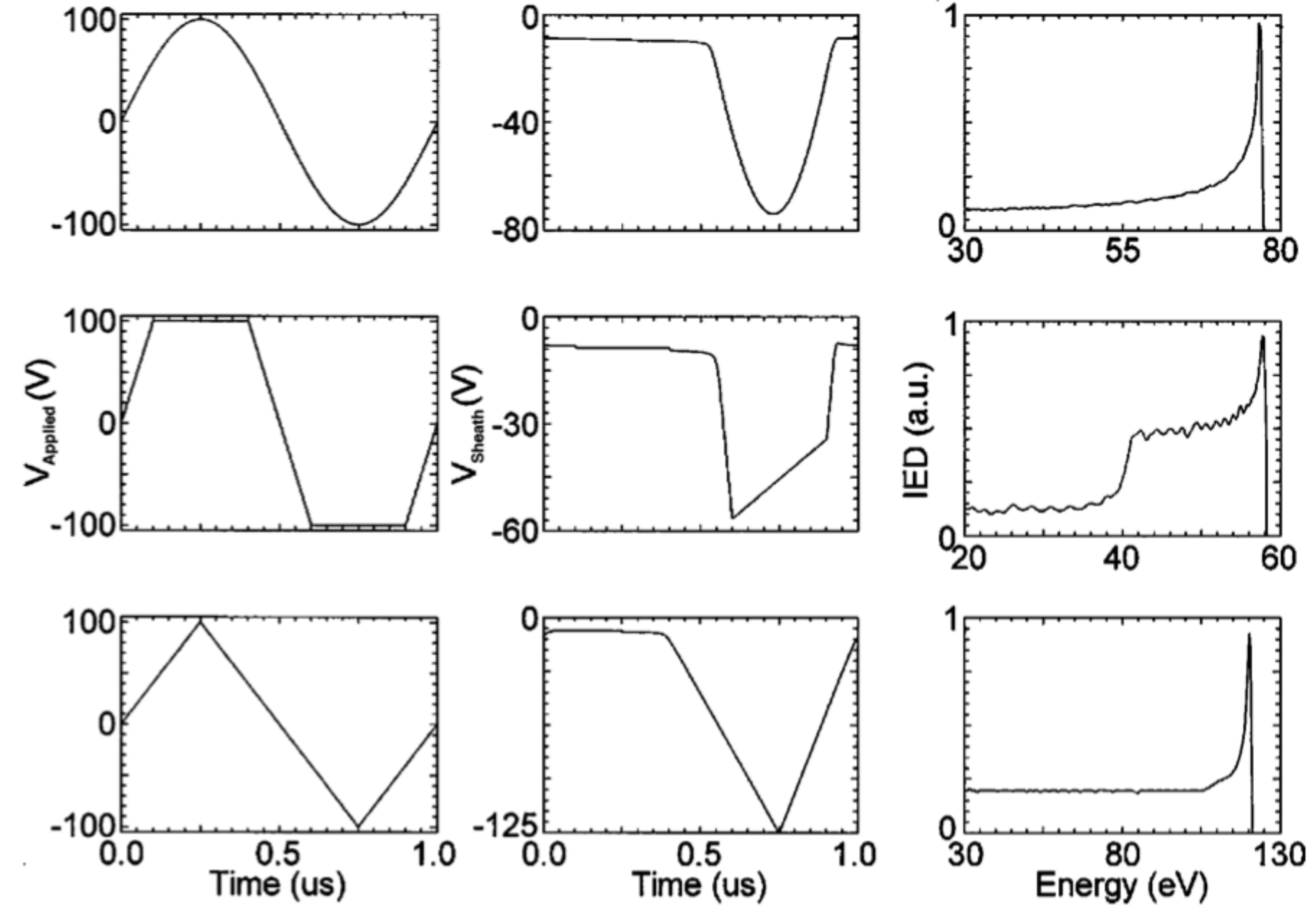}
\caption{Applied radio-frequency bias voltage (left column),
sheath voltage (center collumn), and IEDF (right column) for
sinusoidal (above row), square (center row), and triangular
waveforms (below row) from Rauf \cite{Rauf2000}. The results
has been obtained for inductively coupled plasma in Argon at
20 mTorr with substrate bias voltage of 100 V at 1 MHz.}
\end{figure}

As mentioned in the previous chapters, it is the sheath voltage
which mainly determines the IEDF. The sheath voltage itself can
be adjusted using a bias voltage at the substrate. In this context
Rauf explored the influence of the waveform and frequency of a
radio-frequency bias voltage on the IEDF \cite{Rauf2000}.
Using the HPEM, Rauf found -- as expected -- that the waveform adjusts the
shape of the IEDF while the frequency determines its breadth.
He performed numerical experiments by varying the waveform of the
bias voltage. In figure 8 the results for three different
waveforms are displayed. It can be observed that a sinusoidal
waveform leads to an IEDF that peaks at high energies and gradually
decreases with decreasing energy. A square waveform results in a
sharp step in the IEDF at high energies. The triangular waveform
generates a constant IEDF over a large range of energy.

Rauf also proposed in his paper that one can design
voltage waveforms that produce IEDFs with specific
features by utilizing the correlation between them. This idea has been revisited by Wu et al.
\cite{Wu2007}. They developed an analytical model which is used
to predict the IEDF given the sheath voltage. They assume that
the response of the ions $V_i(t)$ is determined by
$V_i(f)= \alpha(f) V_s(f)$, with $V_i(f)$ and $V_s(f)$
being the Fourier transforms of $V_i(t)$ and $V_s(t)$,
respectively. $\alpha(f)$ is the transfer function which is
defined by
\begin{gather}
\alpha(f)=\frac{1}{\left[(cf\tau_i)^p + 1 \right]^{1/p}}
\end{gather}
The transfer function is chosen in way that the ions have
a $(f\tau_i)^{-1}$ dependence at high frequencies and
completely respond to low frequency oscillations. The constants
$c$ and $p$ are fitting parameters. Once the ion response voltage
$V_i(t)$ is calculated, it will be converted into the corresponding
IEDF using
\begin{gather}
f(E) \propto \sum_j \left| \frac{dV_i}{dt_j}\right|^{-1},
\end{gather}
with $j$ being the energy interval. Figure 9 shows an analytically
calculated IEDF compared to results from a PIC/MC simulation
for an argon discharge driven by two radio-frequency voltages
(800 V at 2 MHz and 400 V at 64 MHz). The analytical IEDF is
shown as the solid line, and the PIC/MC result is shown as the
dashed line. One can observe that for the dual-frequency excitation
of the discharge the	analytical model is able to predict the IEDF,
including its substructures. Calculations of IEDFs performed for
other dual- and triple frequency cases show similarly good agreements.

\begin{figure}
\includegraphics[width=0.7\linewidth]{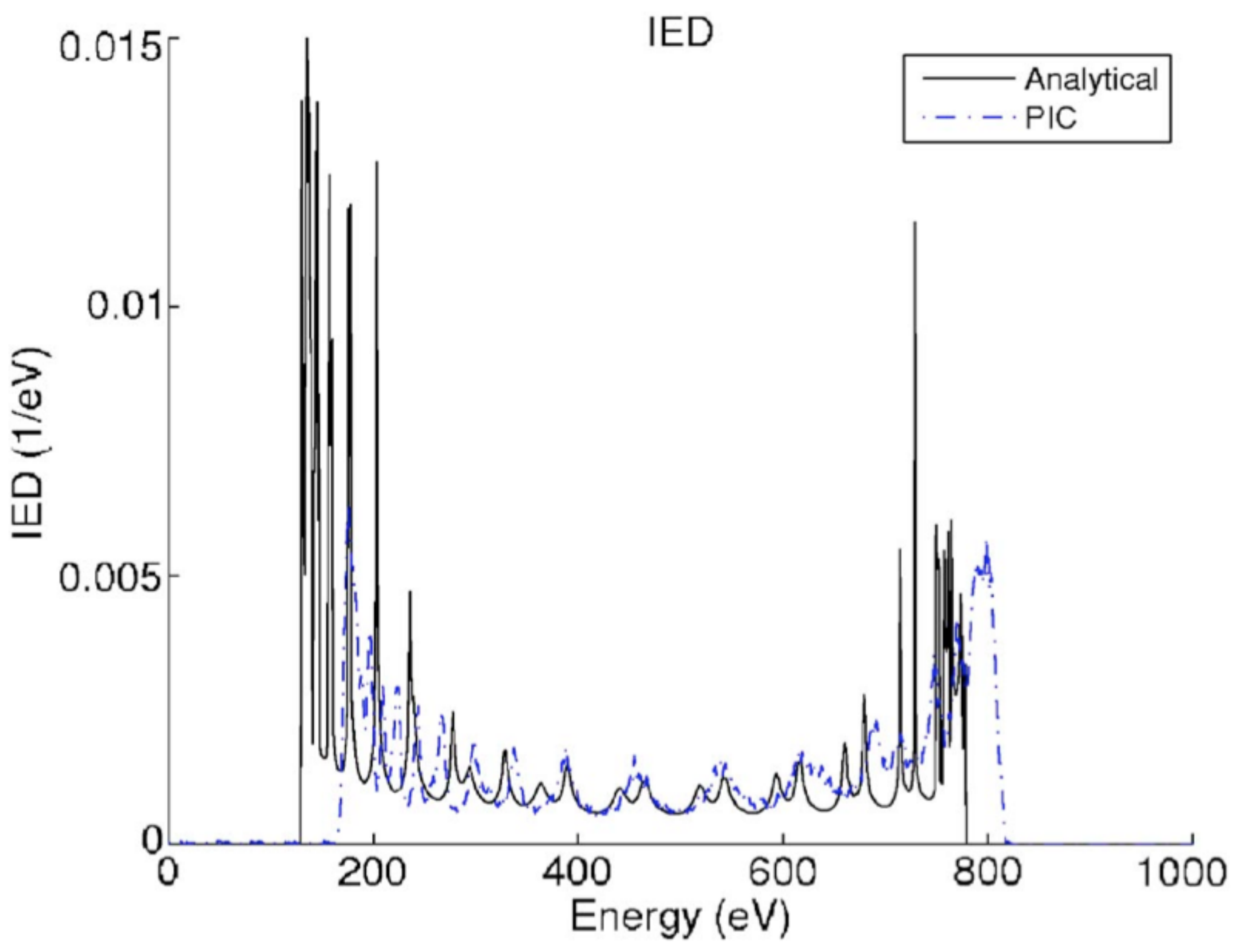}
\caption{IEDF calculated by the analytical model compared to results
from a PIC/MC simulation for an argon discharge driven by
two radio-frequency voltages (800 V at 2 MHz and 400 V at 64 MHz)
from Wu et al \cite{Wu2007}.}
\end{figure}

Two (or even more) distinct radio-frequencies are not only used for
biasing the substrate. Particularly two radio-frequency voltages
have been successfully introduced in order to power the discharge
itself. The so-called dual-frequency CCP concept helps to remedy
the significant drawback of conventional single-frequency CCPs
that the ion bombardment energy and the ion flux to the substrate
cannot be controlled independently. Goto et al. \cite{Goto1992}
as well as Nakano and Makabe \cite{Makabe1995} found that under certain
conditions dual-frequency
CCPs allow for an almost independent control of the ion flux by
the high frequency power and the ion bombardment energy by the
low frequency power. In principle, the mechanism is quite simple:
The source operating at the high frequency ``sees'' the sheath
capacitances as a shortcut; its voltage drops across the bulk,
where it leads to Ohmic dissipation, electron heating, and
ionization. The low frequency voltage, on the other hand, drops
mostly across the sheath; its peak value determines the sheath
voltage and thus the ion energy. For a more detailed discussion
see Lee et al. \cite{Lee2004, Lee2005}, who conducted numerical
investigations on the control of the IEDF in dual-frequency CCPs.
In 2f-CCPs, the ratio of the applied frequencies
$\omega_{\it HF}/\omega_{\it LF}$ is obviously an
important control parameter. By means of simulations using HPEM,
Rauf and Kushner found that at large ratios
$\omega_{\it HF}/\omega_{\it LF}\gg10$
the spectrum of the radio-frequency current through the discharge
is just the superposition of two “single frequency” spectra 
\cite{Kushner1999}. For more comparable frequencies
($\omega_{\it HF}/\omega_{\it LF}\lesssim 10$, however, quite
surprising nonlinear effects were observed, which are indicated
 by the appearance of harmonics and sidebands in the
spectrum \cite{Mussenbrock2006, Ziegler2009, Ziegler2010}.

\begin{figure}
\includegraphics[width=\linewidth]{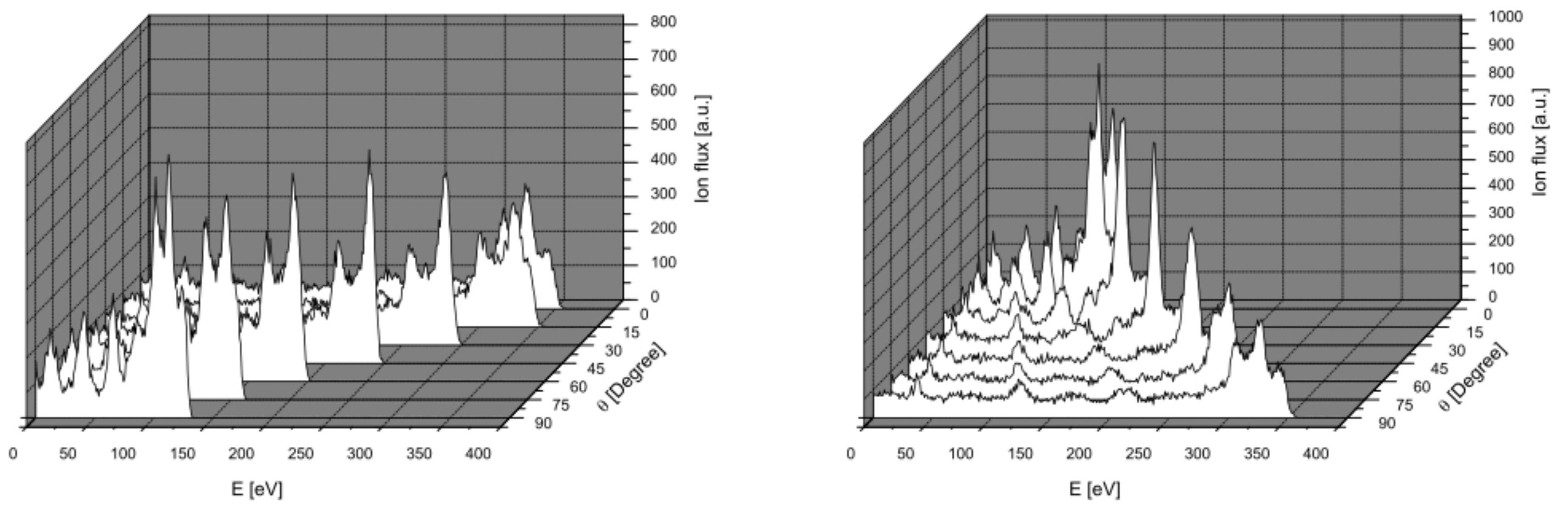}
\caption{Ion flux-energy distributions at the driven (left)
and grounded (right) electrode as a function of the phase
angle $\theta$ obtained by PIC/MCC simulations from Donko
et al. \cite{Donko2009}.
}
\end{figure}
\begin{figure}
\includegraphics[width=0.6\linewidth]{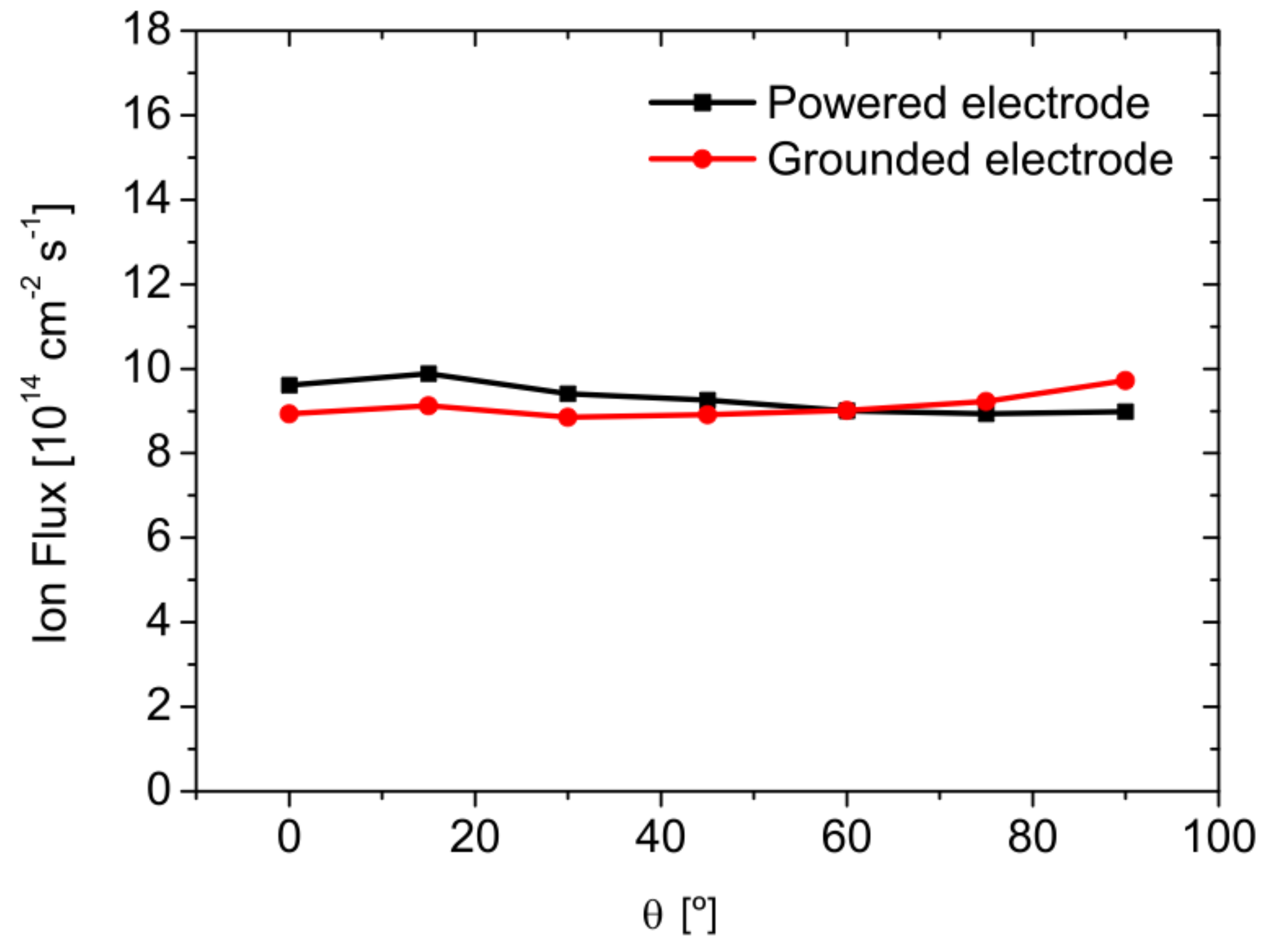}
\caption{Ion fluxes at the driven (left) and grounded (right)
electrode as a function of the phase angle $\theta$ obtained
by PIC/MCC simulations from Donko et al. \cite{Donko2009}[49].
}
\end{figure}

These nonlinear effects can lead under certain conditions to a
strong coupling between the two frequencies, that might limit
separate control of the ion flux and energy. This drawback can be
overcome by making use of the electrical asymmetry effect (EAE).
With the help of the EAE an almost strictly separate control
of the ion energy and flux at the electrode surfaces can be achieved
in a simple way. The EAE can be found in dual-frequency CCPs driven
at comparable frequencies, e.g., 13.56 MHz and 27.12 MHz
\cite{Heil2008a, Heil2008b, Donko2009, Czarnetzki2009}.
If the electrode is driven at a fundamental frequency and its second harmonic
with variable phase shift between the two voltage waveforms,
$V_{\it RF}=\hat V_{\it RF}\left[\cos(\omega t+\theta)+\cos 2 \omega t \right]$,
one obtains a DC self bias voltage as a function of the phase angle $\theta$.
This holds even for geometrically symmetric CCPs were the driven electrode
and the grounded electrode (including grounded walls) are of the same size. This variable
DC self bias adjusted by the phase controls the ion energy at the
electrodes (or the substrate). As only the relative phase between the
voltage harmonics but not their amplitude is changed, the ion flux at
the electrodes remains constant.

\begin{figure}
\includegraphics[width=0.6\linewidth]{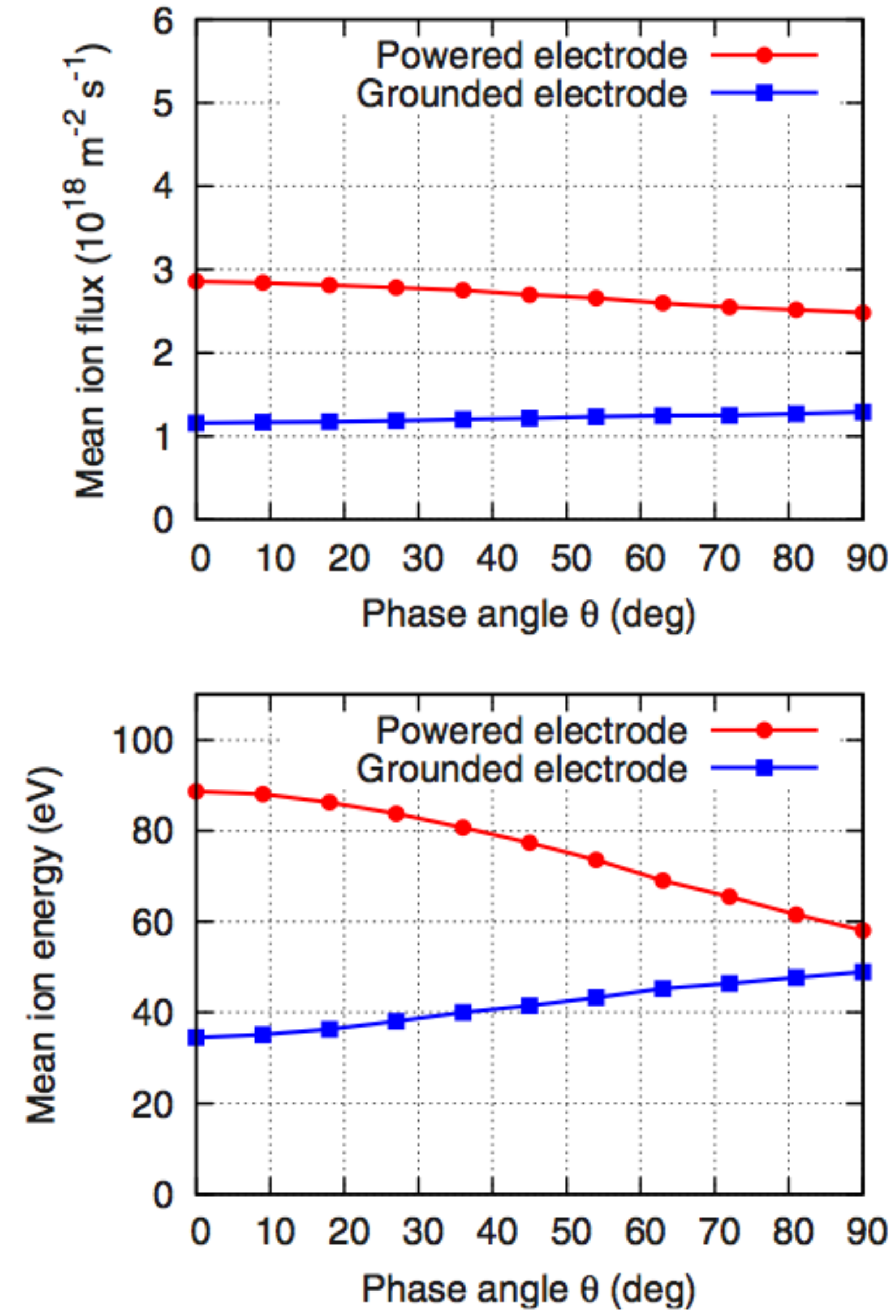}
\caption{Mean flux (top) and mean energy (bottom) of ions at
the smaller powered electrode (red line and dots) and at the
larger grounded electrode (blue line and squares) from Schulze
et al. \cite{Schulze2011}}
\end{figure}

Based on an analytical model developed by Heil et al. \cite{Heil2008a},
one can formulate an expression for the DC self bias voltage $\eta$,
\begin{gather}
\eta = - \frac{{\phi}_{max} + \varepsilon {\phi}_{min}}{1 + \varepsilon}.
\label{DCBias}
\end{gather}
Here, ${\phi}_{max}$ and ${\phi}_{min}$ is the absolute maximum and
absolute minimum of the applied driving voltage waveform, respectively.
$\varepsilon$ is the so-called symmetry parameter defined by
\begin{gather}
\varepsilon = \left| \frac{{\phi}_{sg}}{{\phi}_{sp}} \right| \approx
\left(\frac{A_{p}}{A_{g}} \right)^2 \frac{\bar{n}_{sp}}{\bar{n}_{sg}}
\left(\frac{Q_{mg}}{Q_{mp}} \right)^2.
\label{DefSymPar}
\end{gather}  
${\phi}_{sg}$, ${\phi}_{sp}$ is the maximum voltage drop across the
sheath at the grounded and powered electrode, respectively,
$\bar{n}_{sp}$, $\bar{n}_{sg}$ is the spatially averaged ion
density, and $Q_{mg}$, $Q_{mp}$ is the maximum charge in the
respective sheath. The remarkable feature of the EAE is that
the DC self bias is an almost linear function of the phase angle
between the driving frequencies in range between 0 and $\pi/2$.
This has been confirmed my PIC/MC simulation and fluid models.
Therefore the phase angle $\theta$ is knob for tuning the IEDF.
Donko et al performed 1D PIC/MC simulations of an geometrically
symmetric argon discharge driven at 13.56 MHz and 27.12 MHz
\cite{Donko2009}. Figure 10 shows the obtained IEDFs in front of both
the driven electrode (right) and the grounded electrode (left). One
can observe the effect of varying the phase angle $\theta$ on the
IEDF at both electrodes. By changing $\theta$ from 0 to $\pi/2$
the maximum ion energy at both electrodes can be changed by a
factor of almost 3. Furthermore, the role of each electrode can
be reversed by changing $\theta$. The high ion energy at the
driven electrode at $\theta=0$ can be switched to the grounded
electrode at $\theta=\pi/2$, and vice versa. Another important feature of this behavior is the fact that
the switching from high to low ion energy (and from low to high ion
energy) can be realized under almost constant ion flux. Figure 11  shows 
the ion flux at both electrodes as the
phase angle $\theta$ is varied from 0 to $\pi/2$. The ion flux remains
constant within $\pm$ 5\%, while the maximum ion energy changes by a
factor of 3 as $\theta$ changes. The observed stability
of the ion flux is within the range of tolerance for most industrial
applications.

The EAE can also be used to reduce the asymmetry of geometrically
asymmetric discharges. By tuning the phase angle shift between two
consecutive driving harmonics, the absolute value of the DC self-bias
voltage can be substantially reduced and the mean ion energies at both
electrodes can be made similar. The related results from PIC/MC simulations
are displayed in figure 12. It could be very interesting to work out whether
this effect similarly appears in discharges in molecular, electronegative
gases which are used for materials processing. This might allow to switch
electrically from an etching mode to a cleaning or even a deposition mode
in one and the same reactor. One would avoid the necessity for a change
of the gas mixture for chemical wall cleaning and different chamber
geometries for etching and deposition processes.

\section{Conclusion}

The fundamental concepts of modeling and simulation of IEDFs
from simplified models to self-consistent plasma simulations
are summarized. Finally well-established and new concepts for
controlling the IEDF are discussed. Although the overall number
of contributions to the field of IEDF modeling and simulation
is tremendous and research has been done for more than 40 years
it is still a topic of growing interest. This holds
particularly for the control of IEDFs and even the closed-loop
control and tailoring IEDFs pioneered by Wendt and co-workers
\cite{Wang2000, Wang2001, Silapunt2004, Buzzi2009} and
improved by Baloniak and von Keudell \cite{Baloniak2010a,
Baloniak2010b, Baloniak2010c}
in order to realize specific properties of substrate materials. 
Unfortunately, this exciting topic is beyond the scope of this
paper. However, the basis of all concepts is adapted modeling
of IEDFs.

\section{Acknowledgement}
The author gratefully acknowledges numerous discussions with
Ralf Peter Brinkmann and the people of the Institute of 
Theoretical Electrical Engineering at the Ruhr University Bochum.
The work is supported by the Deutsche Forschungsgemeinschaft DFG
in the frame of SFB-TR 87.


\end{document}